\documentclass[conference]{IEEEtran}
\IEEEoverridecommandlockouts
\usepackage{cite}
\usepackage{amsmath,amssymb,amsfonts}
\usepackage{algorithmic}
\usepackage{graphicx}
\usepackage{textcomp}
\usepackage{xcolor}
\usepackage{comment}
\usepackage{url}
\usepackage{booktabs}
\usepackage{float} 
\usepackage{algorithm}
\usepackage{algorithmic}
\usepackage[caption=false, font=footnotesize]{subfig}
\def\BibTeX{{\rm B\kern-.05em{\sc i\kern-.025em b}\kern-.08em
    T\kern-.1667em\lower.7ex\hbox{E}\kern-.125emX}}

\begin{document}

\title{ Audio Signal Processing Using Time domain Mel Frequency wavelet coefficient\\}
\author{\IEEEauthorblockN{1\textsuperscript{st} Rinku Sebastian}
\IEEEauthorblockA{ \textit{University of York.}\\
York, United Kingdom. \\
rinku.sebastian@york.ac.uk}
\and
\IEEEauthorblockN{2\textsuperscript{nd} Simon O'Keefe}
\IEEEauthorblockA{\textit{University of York.}\\
York, United Kingdom. \\
simon.okeefe@york.ac.uk}
\and
\IEEEauthorblockN{3\textsuperscript{rd} Martin A. Trefzer}
\IEEEauthorblockA{\textit{University of York.}\\
York, United Kingdom. \\
martin.trefzer@york.ac.uk}
}
\maketitle

\begin{abstract}
Extracting features from the speech is the most critical process in Speech signal processing. Mel Frequency Cepstral Coefficients (MFCC) are the most widely used features in the majority of the speaker and speech recognition applications as the filtering in this feature is similar to the filtering taking place in human ear. But the main drawback of this feature is that  it provides only the frequency information of signal but does not provide the information about at what time which frequency is present.
The Wavelet Transform, with its flexible time-frequency window, provides time and frequency information of the signal, is an appropriate tool for the analysis of non stationary signals like speech. On the other hand, because of its uniform frequency scaling, a typical wavelet transform may be less effective in analyzing speech signals, have poorer frequency resolution in low frequencies, and be less in line with human auditory perception. Hence it is necessary to develop a feature that incorporates the merit of both MFCC and Wavelet transform. A great deal of studies are trying to combine both theses features. The present Wavelet Transform based Mel-scaled features extraction methods require more computation when a wavelet transform is applied on top of mel-scale filtering, since it adds extra processing steps. Here we are proposing a method to extract Mel scale features in time domain combining the concept of wavelet transform thus reducing the computational burden of time-frequency conversion and complexity of wavelet extraction. Combining our proposed Time domain Mel frequency Wavelet Coefficient(TMFWC) technique with the reservoir computing methodology has significantly improved the efficiency of audio signal processing.
\end{abstract}

\section{Introduction}
Even with the development of cutting-edge technologies, audio signal processing remains difficult and lacks the precision of a human speech processing system. Several researches merged Wavelet with MFCC to generate MFCC based on Wavelet in an attempt to build a superior feature extraction approach. The benefits of both approaches are combined when they are applied. Although MFCC is based on the paradigm of human auditory perception and may compactly describe the speech spectrum, the frequency domain transformation procedure may result in information loss as well as loss of time knowledge. Wavelet transform was considered an alternative to this problem as it  can translate the signal into the frequency and time domain thus providing both frequency and time information. But for audio signal processing wavelet transform shows poor frequency resolution in low frequencies, less human auditory perception alignment, and potentially less effectiveness in analyzing speech signals. Compared to conventional MFCC and wavelet transform, the Wavelet-MFCC combo produced superior outcomes. Hence researchers are working to create a technique for audio processing, which combines the benefits of wavelet and MFCC. The Mel Frequency Cepstral Coefficient  scale, which mimics human hearing perception, with the time-frequency resolution capabilities of the wavelet transform, allows for better analysis of transient sounds and noise variations within a speech signal. A Mel Frequency Wavelet Transform (MFWT) is specifically designed to mimic the non-linear frequency perception of the human hearing system, which means it provides better detail in the low-frequency range where most speech energy resides.

A "mel frequency wavelet coefficient" refers to a feature extracted from an audio signal by applying a wavelet transform to the signal's spectrum after it has been scaled using the mel frequency scale, essentially capturing both time and frequency information with a focus on human perception of pitch, making it a valuable tool in speech related applications.
The "mel" part indicates that the frequency spectrum is mapped onto a mel scale, which approximates how humans perceive pitch, where lower frequencies are spaced closer together and higher frequencies are spaced farther apart. The "wavelet" part signifies that a wavelet transform is applied to the mel-scaled spectrum, allowing for time-localized analysis of the signal's frequency components. By calculating these coefficients, we obtain a set of features that can be used to characterize the signal, particularly useful for tasks like speaker recognition and sound classification. 
\begin{figure}
    \centering
    \includegraphics[width=.5\textwidth]{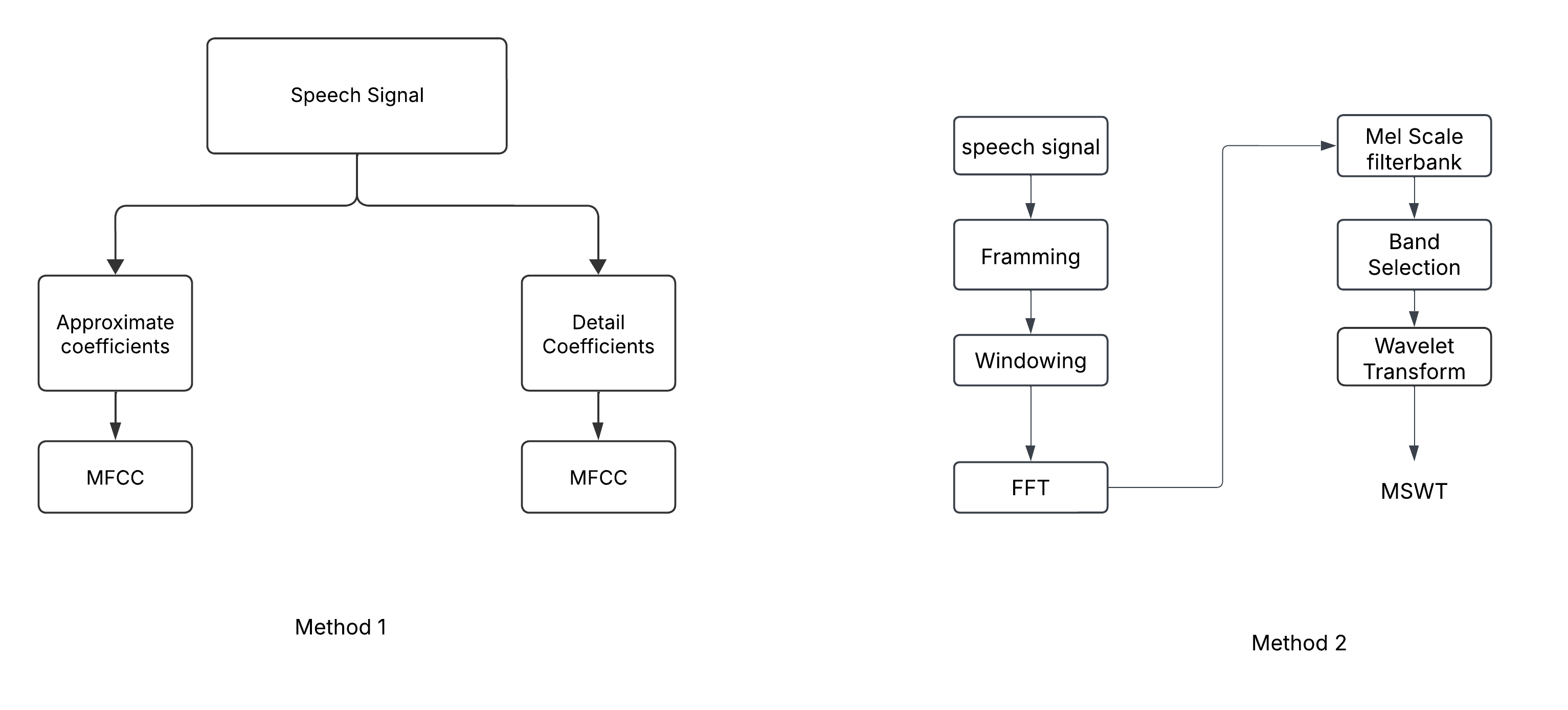}
    \caption{wavelet based mel frequency coefficient extraction methods}
    \label{fig:enter-label 1}
\end{figure}

In the state of art method of extraction of wavelet based mel frequency coefficient, wavelet transform is done prior to MFCC part or after MFCC part as shown in figure 1. In either case the whole step of calculation of both MFCC and wavelet transform is done to obtain the  Wavelet Transform based Mel-scaled Feature extraction. This makes the whole process complicated. In our approach, we use the time domain feature extraction method to extract the mel frequency wavelet coefficient, reducing the method's complexity and increasing its efficiency. The time-domain capability of a reservoir computing technique is also made use to improve the performance of the entire system.

\section{ Reservoir computing}

Reservoir computing is a bio-inspired paradigm in machine-learning. It is a framework for computation that was developed from the notion of recurrent neural networks that maps input signals into higher dimensional computational spaces via the dynamics of a fixed, non-linear system known as a reservoir. After the input signal is fed into the reservoir, which is treated as a `black box', a straightforward readout mechanism is trained to read the state of the reservoir and map it to the desired output. An RNN is created at random and it is just the readout which trained in reservoir computing, typically using some regression based on least squares.
\begin{figure}
    \centering
    \includegraphics[width=\columnwidth]{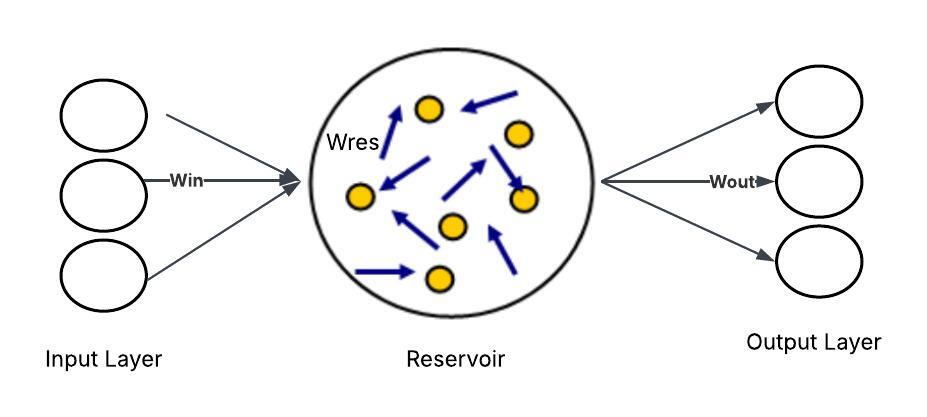}
    \caption{Topology of Reservoir computer}
    \label{fig:enter-labelr1}
\end{figure}

Since RNN development is sluggish and challenging, in 2001 Wolfgang Maass and Herbert Jaeger independently suggested Liquid State Machines~\cite{Maass2002-tg} and Echo State Networks~\cite{Jaeger_2010-pc} as fundamentally new approaches to RNN design and training. Reservoir Computing is a term that has since been coined to refer to these methods. It has roots in computational neuroscience~\cite{Dominey2004ComplexSS} and later consequences in machine learning as the Backpropagation-Decorrelation~\cite{Steil2004-kk} learning rule (RC).  Figure~\ref{fig:enter-labelr1} shows a classical reservoir computer. An input layer that is randomly connected to each of the N reservoir nodes receives the input. The reservoir itself is left untrained since the connections and weights between its nodes are fixed and selected at random.  An output layer reads out the transient dynamical response of the reservoir using linear weighted summing of the node states. The drawbacks of gradient-descent RNN training are avoided by the RC paradigm. This made it much easier to use RNNs in real-world applications and outperformed traditional fully trained RNNs in many tasks~\cite{Lukosevicius2009-ib}.

In the reservoir framework, since the training is limited to the readout part, the burden of training is reduced. Also, interference between the tasks is also minimized if we are performing multiple tasks by training multiple readouts on the same reservoir. It is possible to solve several tasks with a single input by adding multiple readouts to a single reservoir. So multitasking can be efficiently or effectively employed using reservoirs. The echo state property of a reservoir gives the system memory so that it can process time series. The fading memory property of reservoir allows the system not to saturate. Furthermore, the reservoir has the  ability to perform nonlinear transformations. All these qualities of a reservoir show that it is a suitable fit for temporal signal processing~\cite{Ghani2010-jr}.

\section{Audio signal processing}

Analyzing an audio signal entails extracting its qualities, forecasting its behaviour, identifying any patterns it may include, and determining how one signal relates to other signals of a similar nature. Music, conversation, and environmental noises are all examples of audio signals. In terms of signal analysis and classification, audio signal processing has developed tremendously over the past decades. Additionally, it has been demonstrated that many current problems can be resolved by combining advanced machine learning (ML) algorithms with audio signal processing methods. Any ML algorithm's performance is based on the features used for training and testing. Consequently, one of the most crucial steps in a machine learning process is feature extraction~\cite{Sharma2020-pb}.

Feature extraction is a method of extracting the dominant and distinctive qualities of a signal. The process of feature extraction involves converting an audio waveform into a parametric representation at a data rate that is relatively low for further processing and analysis. The goal of feature extraction is to represent an audio signal using a fixed number of components. This is due to the fact that processing all of the information in the acoustic signal would be intractable, and some of it is not relevant for the purpose~\cite{Ajibola_2018-ng}. An appropriate feature mimics a signal's characteristics in a much more condensed manner.

The following section describes Mel Frequency Cepstral Coefficient in detail.

\subsection{MFCC}
Mel-frequency Cepstral Coefficients are referred to as MFCC. The Mel-scale used is to map between linear frequency scale of speech signals to logarithmic scale for frequencies higher than 1 kHz. This makes the spectral frequency characteristics of a signal closely corresponding to human  auditory perception and hence,  MFCCs are a feature that is frequently used in automatic speech and speaker recognition. The mel-frequency cepstrum (MFC), which is based on a linear cosine transform of a log power spectrum on a nonlinear Mel scale of frequency, is a representation of the short-term power spectrum of a sound. An MFC is made up of a number of coefficients known as Mel-frequency cepstral coefficients (MFCCs). The frequency bands of the MFC are evenly spaced on the Mel scale.This frequency warping may make it possible to depict sound more accurately.

MFCCs are commonly derived as follows:
\begin{itemize}
    \item 	Step 1: Take the Fourier transform of (a windowed excerpt of) a signal.
\item	Step 2: Map the powers of the spectrum obtained above onto the Mel scale, using triangular overlapping windows or alternatively, cosine overlapping windows.
\item	Step 3: Take the logs of the powers at each of the Mel frequencies.
\item Step 4: Take discrete cosine transform of the list of Mel log powers.
\item	The MFCCs are the amplitudes of the resulting spectrum
\end{itemize}
\begin{figure}
    \centering
    \includegraphics[width=\columnwidth]{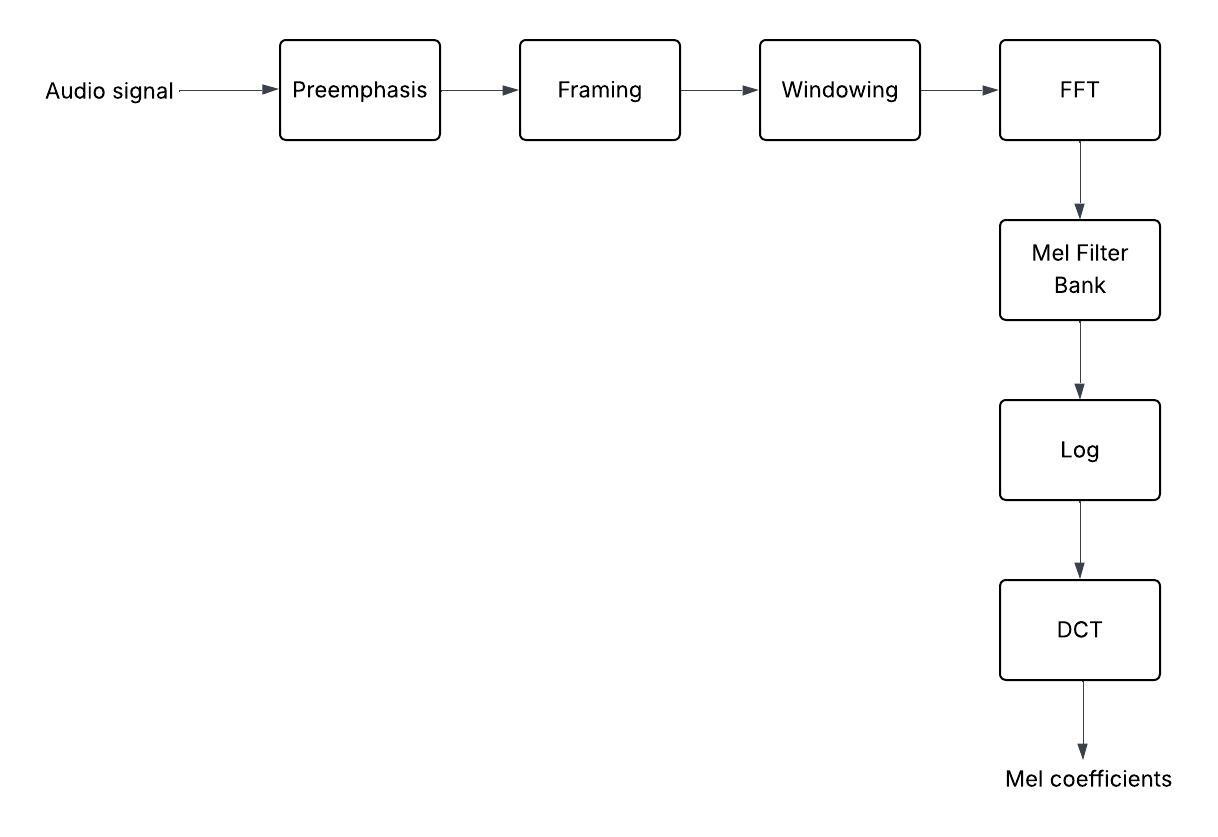}
    \caption{MFC extraction }
    \label{fig:my_label}
\end{figure}
\textbf{Framing and windowing:} The MFCC algorithm needs to be transformed from the time domain to the frequency domain because it is based on spectral analysis. The acoustic signal is essentially stationary. The signal is believed not to be periodic for sound samples that are longer than 200 milliseconds. It is impossible to identify whether a sample that lasts between 30 and 200 milliseconds is periodic or not. It is safe to presume that a sound is periodic for samples that are shorter than 30 ms. ~\cite{Niewiadomy_2008-zy}.  There should be between 20 to 30 milliseconds between each frame. Individual speech sounds' temporal properties can be followed by moving the time window forward by 10 ms at a time, and a 20 ms analysis window is typically long enough to resolve major temporal characteristics while still giving these sounds acceptable spectral resolution. The goal of the overlapping analysis is to ensure that each speech sound in the input sequence is roughly centered within a specific frame. The signal is tapered towards the frame borders on each frame by applying a window. Hanning or Hamming windows are typically used. While applying the discrete Fourier transform (DFT) to the signal, this is done to improve the harmonics, soften the edges, and to reduce the edge effect.

\textbf{DFT spectrum:} Each windowed frame is converted into frequency spectrum by applying DFT.
\begin{equation}
    X(k)=\sum_{n=0}^{N-1} x(n)*e^ {-j2\pi nk/N} 
\end{equation}

\textbf{Mel spectrum:}  Mel spectrum is computed by passing the Fourier transformed signal through a set of band-pass filters known as Mel-filter bank. A Mel is a unit of measurement of how loudness is perceived by the human ear. Since the human auditory system reportedly does not detect pitch linearly, it does not correspond linearly to the tonal frequency physically present in the sound. The frequency spacing for the Mel scale is roughly linear below 1~kHz and logarithmic above 1~kHz. Mel can be approximated by physical frequency using the formula
\begin{equation}
    f_{Mel}=2595log_{10}(1+f/700)
\end{equation}
Where f denotes the physical frequency in Hz, and $f_{Mel}$ denotes the perceived frequency
\begin{figure}
    \centering
    \includegraphics[width=0.5\textwidth]{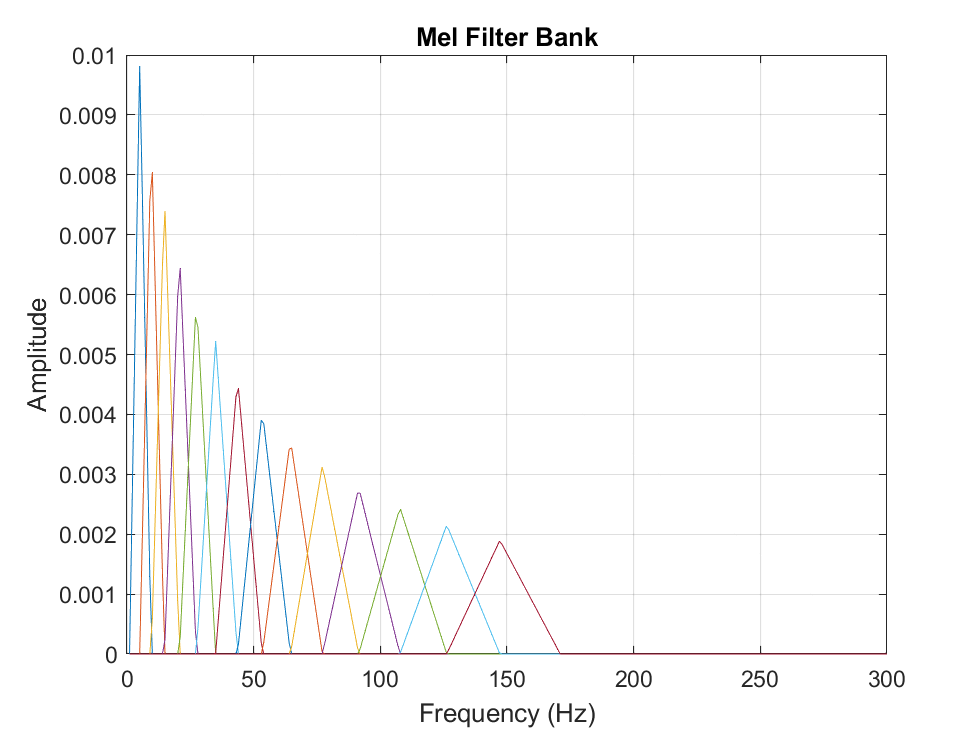}
    \caption{Mel filter bank}
    \label{fig:my_label21}
\end{figure}
Both the frequency domain and the time domain are capable of representing filter banks. Filter banks are typically built in the frequency domain for MFCC calculations. On the frequency axis, the center frequencies of the filters are typically uniformly spaced. However, the warped axis, in accordance with the nonlinear function provided in equation (5), is implemented in order to match the human ear's perception ~\cite{Rao2017-hz}. The filter bank typically consists of overlapping triangular filters~\cite{Molau_undated-vw}. Figure~\ref{fig:my_label21} shows the generated Mel filter bank for 1024 point FFT transform, where the number of filters is 25, minimum frequency is 0 Hz, maximum frequency is 4000 Hz and sampling frequency is 8 kHz. The algorithm generating MFCCs  creates the filter bank before processing is done, because filter bank parameters are constant. The frequency spectrum of the signal(i.e., X(k) from equation (4) is multiplied with the filter bank to obtain mel frequency spectrum. Thus mapping the power-spectrum of the signal on to the Mel scale.  

\textbf{Discrete cosine transform (DCT):} The vocal tract is smooth and hence there is a tendency for adjacent bands' energy levels to correlate. The DCT is used to create a set of cepstral coefficients from the transformed Mel frequency coefficients. The Mel spectrum is typically displayed on a log scale before being subjected to DCT. In the cepstral domain, this produces a signal with a quefrency peak that corresponds to the signal's pitch and a number of formants that represent low quefrency peaks. Since the first few MFCC coefficients constitute the majority of the signal information, the system can be made robust by extracting only those coefficients while ignoring or truncating higher-order DCT components.

Finally, MFCC is calculated as
\begin{equation}
    c(n)=\sum_{m=0}^{M-1} log_{10} (s(m))cos(\pi n(m-0.5)/M)
\end{equation}
n=0, 1,2....C-1.
where c(n) are the cepstral coefficients, and C is the number of MFCCs. MFCC systems use only 8–13 cepstral coefficients. The zeroth coefficient is often excluded since it represents the average log-energy of the input signal, which only carries small amount of speaker-specific information. \cite{Rao2017-hz}

The log Mel spectrum is converted back to the time domain in this last phase, resulting in the MFCCs. For the specified frame analysis, the cepstral representation of the speech spectrum gives a good representation of the local spectral features of the signal. The discrete cosine transform can be used to translate the Mel spectrum coefficients into the time domain because they are real numbers, as is their logarithm (DCT). The log Mel spectrum is transformed back to time in this final stage. The Mel Frequency Cepstrum Coefficients are the outcome (MFCC). The Mel coefficients are transformed back into the time domain using the discrete cosine transform~\cite{Tiwari_2010-xe}.

\textbf{Deltas and Delta-Deltas:} Deltas and Delta-Deltas are also known as differential and acceleration coefficients. Only the power spectral envelope of a single frame is described by the MFCC feature vector, but it would seem that speech would also contain information about dynamics, i.e., the trajectory of the MFCC coefficients over time. It turns out that adding the MFCC trajectories to the original feature vector after computing them, significantly improves automatic speech recognition performance. The benefit of Delta features is that they are used to represent the temporal information. To calculate the delta coefficients, the following formula is used.
\begin{equation}
    d_t = \frac{\sum_{n=1}^N n (c_{t+n} - c_{t-n})}{2 \sum_{n=1}^N n^2}
\end{equation}
where $d_t$ is a delta coefficient from frame t computed in terms of the static coefficients $c_{t-n}$ to $c_{t+n}$. n is usually taken to be 2. By taking the derivative of Delta features, Delta-Delta features are extracted~\cite{Singh_2016-vk}.

\subsection{Wavelet Transform}
\begin{figure}
    \centering
    \includegraphics[width=0.8\linewidth]{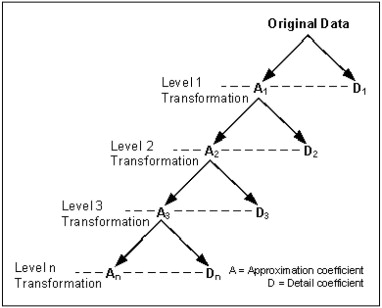}
    \caption{Wavelet decomposition\cite{Russell2008-qh}}
    \label{fig:enter-label5}
\end{figure}

The decomposition of a signal into a collection of basis functions made up of contractions, expansions, and translations of a mother wavelet function $\psi(t)$, referred to as the wavelet transform (WT). The wavelet reduction method is based on the multi-resolution signal decomposition method developed by \cite{Mallat1989674}.The wavelet transform is an efficient noise reduction technique. It is employed to decompose a signal into scaled and shifted representations of specific wavelets.  There are wavelet families that can be used. Two filters are used in the decomposition process, convolving the input signal and subsequently decimating it into detail coefficients (high frequency component) and approximation coefficients (low frequency component). The procedure is carried out repeatedly until a final level is attained. At each level, the approximation coefficient is used to decompose the original data n times. A graphic representation of the breakdown process at each stage is provided in figure \ref{fig:enter-label5}.

The Wavelet transform technique is employed for both temporal and frequency domain analysis. At different frequency bands, the original signal is divided into a large number of components. The wavelet transform of a signal x(t) is defined as:
\begin{equation}
    X(a, b) = \frac{1}{\sqrt{a}} \int_{-\infty}^{+\infty} x(t) \psi \left( \frac{t-b}{a} \right) dt
    \label{eq:wavelet_transform}
\end{equation}

\section{Methodology}
MFCC is generally obtained as shown in Figure~\ref{fig:my_label}. The reason to do that is to simplify the computation. Convolution in time domain is equivalent to multiplication in frequency domain and is equivalent to addition in log-frequency domain. 
\begin{figure}
    \centering
    \includegraphics[width=.52\textwidth]{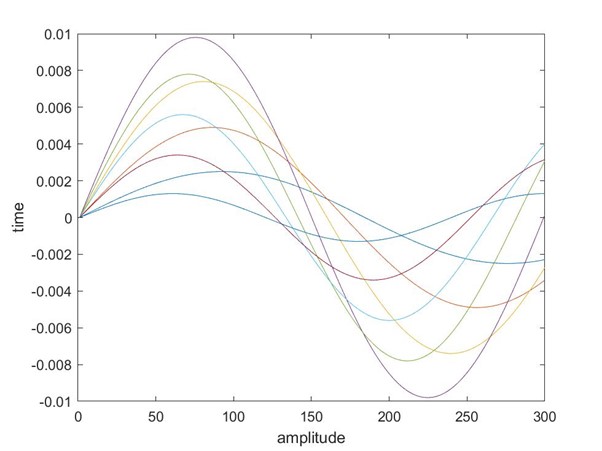}
    \caption{synthesized sine wave corresponding to 1 mel coefficient}
    \label{fig:_FB1}
\end{figure}

\begin{figure}
   % \centering
    \includegraphics[width=.52\textwidth]{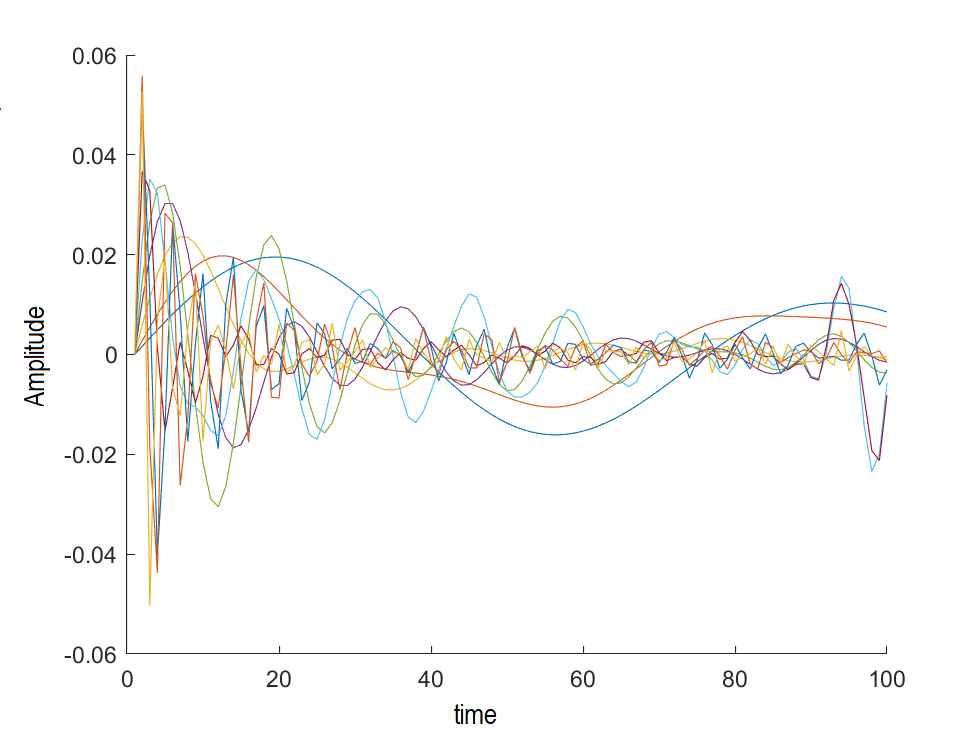}
    \caption{Time-domain filterbank}
    \label{fig:_FB}
\end{figure}

We have created the time domain filter bank signal corresponding to each Mel coefficient. For each Mel coefficient there is a set of frequencies and a set of parameters as shown in Table~\ref{tab:my_label}, where the frequency and parameters for first Mel coefficient are given. We have synthesized a sine wave and cosine wave corresponding to each frequency and parameter as shown in figure \ref{fig:_FB1}. We then superimposed all the synthesized sine waves to get imaginary part of wavelet transform in terms of  Mel filter bank signal. Similarly superimposed all the synthesized cosine waves to get real part of wavelet transform in terms of  Mel filter bank signal. Thus synthesized all the Mel filter bank signals. Figure \ref{fig:_FB} shows the Mel filterbank in time domain.

\begin{table}
    \centering
    \begin{tabular}{@{}cccccccccc@{}}
    \toprule
         parameter= & 0.002454697 & 0.004909393 & 0.00736409& 0.009818787 \\
         %\hline
         $\Delta f$= & 131 & 141 & 151 & 161 \\
         \midrule
         parameter= &0.007781114 & 0.00561907 & 0.003457026 & 0.001294982\\
         %\hline
         $\Delta f$= & 171 & 181 & 191 & 201 \\
         %\hline
         \bottomrule
    \end{tabular}
    \vspace{3px}
    \caption{Parameters and frequencies for 10 Mel frequencies used}
    \label{tab:my_label}
\end{table}

In order to obtain MFWC in the time domain, the audio signal is convoluted separately with sine and cosine mel-wave corresponding to each of the Mel filterbank signals. 

The "magnitude of a wavelet transform" refers to the absolute value of the wavelet coefficients obtained after applying a wavelet transform to a signal, essentially representing the strength or intensity of the signal components at different scales and locations in the time-frequency domain; it indicates how well the signal aligns with the chosen wavelet function at a specific scale and time position. A larger magnitude value in a wavelet coefficient signifies a stronger presence of the corresponding frequency component within the signal at that specific time window. 
 
\begin{figure}
    \centering
    \includegraphics[width=0.8\linewidth]{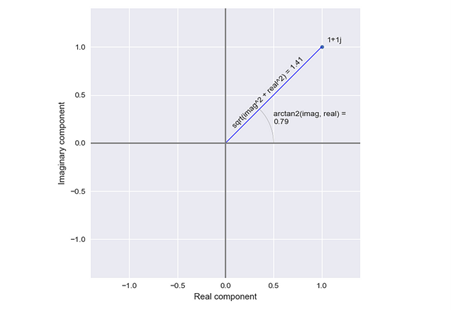}
    \caption{finding magnitude}
    \label{fig:enter-labelM}
\end{figure}

The magnitude of the corresponding signal is obtained using the equation
   $\sqrt{\text{imaginary}^2 + \text{real}^2}$. 
  We have made use of the concept in Figure \ref{fig:enter-labelM} to find the magnitude of the corresponding signal. The signal thus obtained is called Time domain Mel Frequency Wavelet Coefficient (TMFWC). Here the number of data-points of the coefficient  is almost same as the signal. So we have to use some data reduction method before feeding the data to a reservoir for classification. we have used the absolute max-pooling technique  for the same. as in this case the the largest coefficient among the small interval of data corresponds to the coefficient with strongest signal information the data reduction method does not affect the information content of the signal. This signal and is given to the reservoir for classification.

\section{Experiment and Results}
To evaluate the performance of the proposed methods we have used both the Ti-46 data-set and Audio-Mnist dataset for our studies. We have tried to identify the speaker as well as digit using the TMFWC as the feature. We have calculated the percentage of correct utterance across 10 reservoirs and plotted the result. The result is as shown in the figure \ref{fig:enter-labelRD}.
\begin{figure}[h]
    \centering
   \subfloat[Digit Recognition performance]
    {\includegraphics[width=0.55\linewidth]{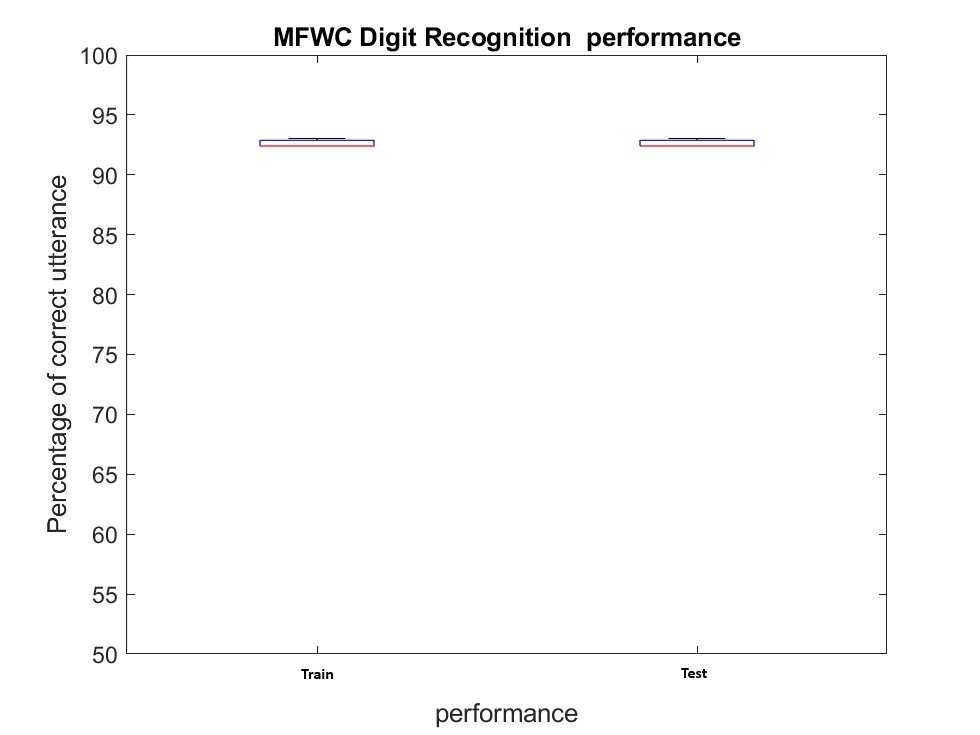}}
    \subfloat[Speaker Recognition performance]
     {\includegraphics[width=0.55\columnwidth]{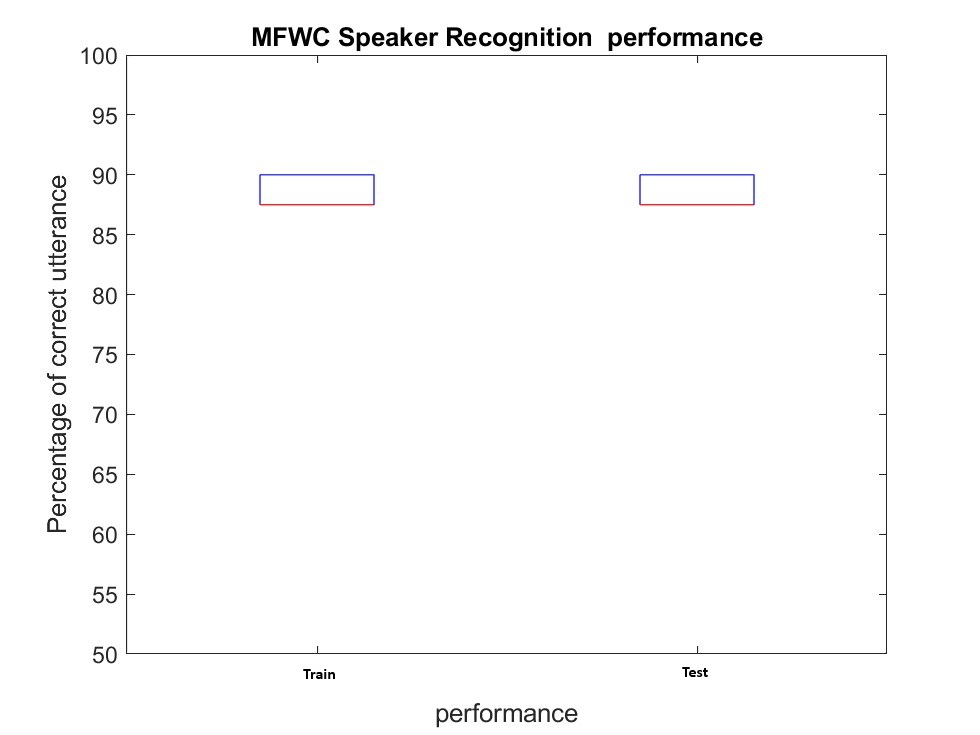}}
            \caption{performance of our system with Ti-46 dataset}
    \label{fig:enter-labelRD}
\end{figure}

\begin{figure}[h]
    \centering
   \subfloat[Digit Recognition performance]
    {\includegraphics[width=0.55\linewidth]{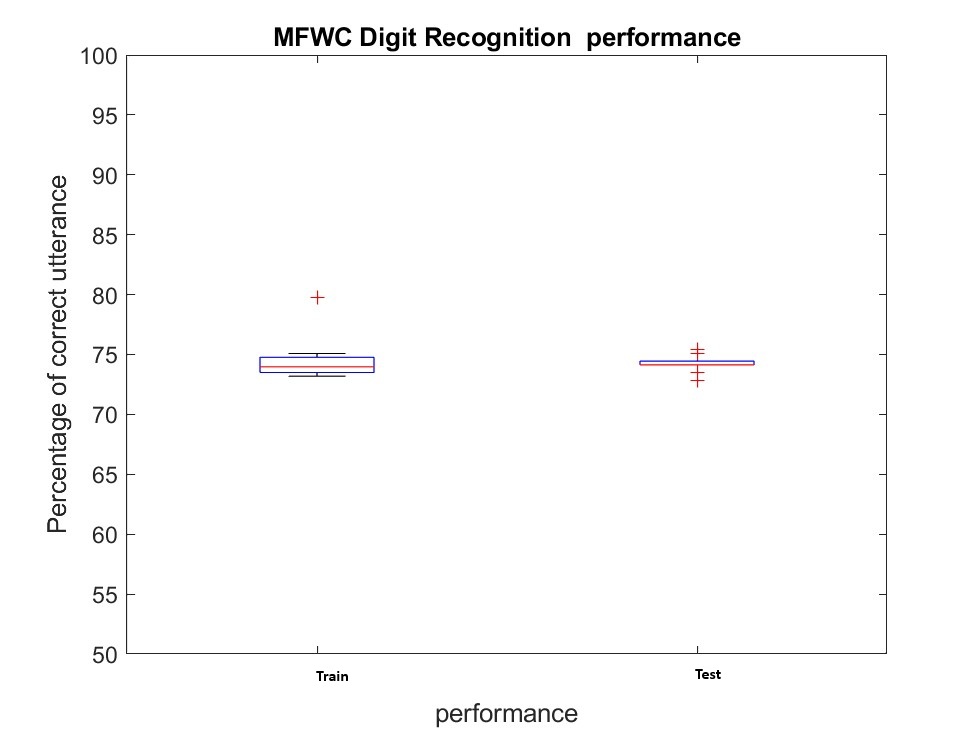}}
    \subfloat[Speaker Recognition performance]
     {\includegraphics[width=0.55\columnwidth]{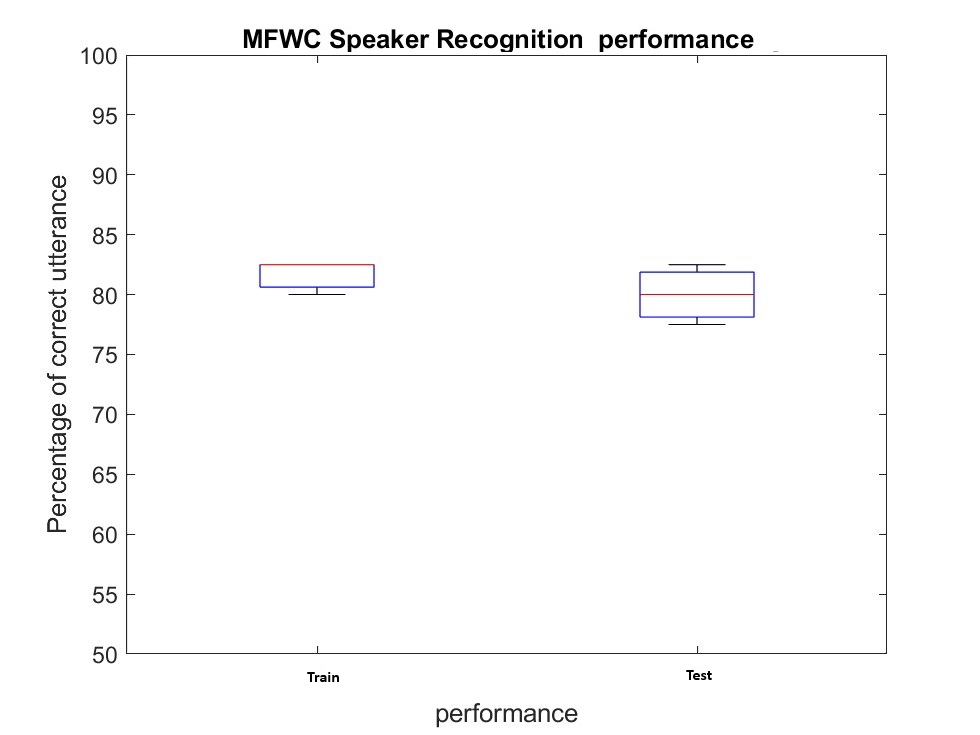}}
            \caption{performance of our system with Audio-Mnist dataset}
    \label{fig:enter-labelRD}
\end{figure}
By demonstrating methods for implementing efficient time-domain Mel Frequency Wavelet Coefficient (MFWC) extraction, we present a framework that improves the efficiency of real-time audio processing. Our approach reduces computational complexity, particularly in performing the Fourier and wavelet transforms, by eliminating the need for complex time-frequency conversion. These results are promising, as the system delivers competitive performance while significantly reducing computational overhead. 

\section{Conclusion}
In this paper, we discussed the usefulness of the time-domain mel frequency wavelets coefficient in speech signal processing.  We presented a novel idea of generating the MFWC in time domain  minimizing the complexity of the existing method. The TMFWC  feature  displayed  more  discriminative  power than other mel-scale based  or wavelet features.  Accompanied by Reservoir as the classifier, the method has significantly improved the efficiency of speech  signal processing.

\bibliographystyle{plain}
\bibliography{Reference}

\end{document}